\documentclass[12pt]{article}
\usepackage{amssymb,amsmath,amsthm}
\usepackage{latexsym}

\newcommand{\beg}{\begin{equation}}
\newcommand{\ene}{\end{equation}}

\begin{document}
\title{
\textsc{\bf Towards Quantum Information Theory in } \\
\textsc{\bf Space and Time}
\\
$~$\\}
\author{
\textsf{Igor V. Volovich}
\\
\emph{Steklov Mathematical Institute}\\
\emph {Russian Academy of Sciences}\\
\emph{Gubkin St. 8, 119991, GSP-1, Moscow, Russia}\\
\emph{e-mail: volovich@mi.ras.ru}
}

\date {~}
\maketitle
\thispagestyle{empty}

\begin{abstract}
Modern quantum information theory deals with an idealized situation  when
the spacetime dependence of quantum phenomena is neglected. However the
transmission and processing of (quantum)  information is a physical process
in spacetime. Therefore such basic notions in quantum information theory as
qubit, channel,   composite systems and entangled states should
be formulated in space and time. In particlular we suggest
that instead of a two level system (qubit) 
the basic notion in a relativistic quantum information theory
should be a notion of an elementary
quantum system, i.e.
an infinite dimensional Hilbert space $H$
invariant under an irreducible representation of the Poincare group
labeled by $[m,s]$ where $m\geq 0$ is  mass and $s=0,1/2,1,...$
is  spin.
We emphasize an importance of
consideration of quantum information theory from the point of view of
quantum field theory. We point out and discuss   a fundamental fact that in
quantum field theory there is a statistical dependence  between two regions
in spacetime even if they  are spacelike separated. A classical
probabilistic representation for  a family of correlation functions in
quantum field theory is obtained. Entangled states   in space and time are
considered. It is  shown that any reasonable state in  relativistic quantum
field theory  becomes disentangled (factorizable) at large spacelike
distances if one makes local observations. 
As a result a  violation of Bell`s  inequalities can be observed
without inconsistency with principles of relativistic quantum theory only
if the distance between detectors is rather small.
We suggest a further experimental study of entangled states 
in spacetime by studying
the dependence of the correlation functions on the distance
between detectors.
\end{abstract}
\newpage

\tableofcontents

\section{Introduction}
\label{intro}

Remarkable experimental and theoretical results 
obtained in quantum computing, teleportation and cryptography
(these topics sometimes are considered as
belonging to quantum information theory) 
are based on the  investigation of basic properties of quantum
mechanics. Especially important are  properties of nonfactorized
entangled states introduced by Einstein, Podolsky and Rosen
which were named by Schrodinger as the most characteristic 
feature of quantum mechanics.

 Ideas of Shannon`s
classical information theory  are important
for the modern quantum information theory
as well as 
the notions of qubit, quantum relative entropy,   quantum channel,  
and  entangled states , see for example
~\cite{QI} - \cite{Hol}.

The spacetime dependence is not explicitly indicated
in this approach. As a result,  
many important achievements  in modern quantum information theory
have been obtained for an idealized situation 
when the spacetime dependence
of quantum phenomena is neglected. 

We emphasize the importance of 
the investigation of quantum information effects in space and time.
\footnote{The importance of 
the investigation of quantum information effects in space and time
and especially the role
of relativistic invariance in classical and quantum information theory
was stressed in the talk by the author  
at the First International Conference
on Quantum Information which was held at Meijo University, Japan,
November 4-8, 1997.}
Transmission and processing of (quantum) information is a physical process
in spacetime.  Therefore a formulation of 
such basic notions in quantum information theory
as  composite systems, entangled states and the 
channel should include 
the spacetime variables \cite{Vol3}.

Ultimately, quantum information theory should become
 a part  of quantum field theory (perhaps, in future, a part
 of superstring theory) since quantum field theory is our most fundamental
 physical theory.
 
Quantum field theory \cite{BSch} is not just an 
abstract mathematical theory 
of operators in a Hilbert space. Basic equations of quantum field theory
such as the Maxwell, Dirac, Yang--Mills equations are differential
equations for operator functions defined on the spacetime.
The nonrelativistic Schrodinger equation is also a 
differential equation
in spacetime.
Therefore a realistic quantum information theory should be based on
the study of the solutions of these equations propagated in spacetime.

One could suggest to define a context described in \cite{Khr2} 
as a boundary condition
for a differential equation. Then we would derive
the contextual dependence
of probabilities   from the study of the dependence
of solutions of the  equation on the boundary conditions.

In modern quantum information theory the basic notion is the two
dimentional Hilbert space, i.e. qubit. We suggest that in a relativistic
quantum information theory, when the existense of spacetime
is taken into account, the basic notion should be a notion of an elementary
quantum system, i.e. according to Wigner (see \cite{Bog}) it is
an infinite dimensional Hilbert space $H$
invariant under an irreducible representation of the Poincare group
labeled by $[m,s]$ where $m\geq 0$ is  mass and $s=0,1/2,1,...$
is  spin (helicity). 

Entangled states, i.e. the states of two particles
with the wave function which is not a product of the wave functions
of single particles,  have been studied in many 
theoretical and experimental
works starting from the paper of Einstein, Podolsky and Rosen,
see e.g.~\cite{AfrSel}.

J. Bell proved ~\cite{Bel} that  there are quantum
spin correlation functions in entangled states
that can not be represented as classical
correlation functions of separated  random variables.
Bell's
theorem reads, see \cite{Vol1}:
$$
\cos(\alpha - \beta)\neq E\xi_{\alpha}\eta_{\beta}
$$
where $\xi_{\alpha}$ and $\eta_{\beta}$ are two random
processes such that $|\xi_{\alpha}|\leq 1$,~
$|\eta_{\beta}|\leq 1$ and $E$ is the expectation.
Here the function $\cos(\alpha - \beta)$ 
describes the quantum mechanical correlation of
spins of two entangled particles. 
Bell`s theorem
has been
interpreted as incompatibility of the requirement of locality with
the statistical predictions of quantum mechanics~\cite{Bel}. For a
recent discussion of Bell's theorem and Bell`s inequalities
see, for example
~\cite{AfrSel} - ~\cite{WW} and references therein.

However
if we want to speak about locality in quantum theory
then we have to localize somehow our particles.
For example we could measure the density
of the energy or the position
of the particles simultaneously with the spin. Only then
we could come to some conclusions about
a relevance of the spin correlation function to the problem
of locality.

 The function
$\cos(\alpha - \beta)$
describes quantum correlations of two spins in the two qubit Hilbert space
when the spacetime dependence of the wave functions of the particles
is neglected.
Let us note however that the very 
formulation of the problem of locality in
quantum mechanics prescribes a special role to the
position in ordinary three-dimensional space. 
It is rather
strange therefore that the problem of {\it local in space
observations} was neglected in   discussions of the problem of locality
in relation to Bell's inequalities . 

Let us stress  that we discuss here not a problem of interpretation of
quantum theory
but a problem of how to make correct quantum mechanical computations
describing an experiment with two  detectors localized in space.
Recently it was  pointed 
out \cite{Vol1} that  if we make  {\it local} observations
of spins then the spacetime part of the wave function leads to an extra 
factor in quantum correlations and as a result the ordinary 
conclusion from the Bell theorem about the nonlocality
of quantum theory fails.

 We present a modification of Bell`s equation which
includes space and time variables.
The function $\cos(\alpha - \beta)$ 
describes the quantum mechanical correlation of
spins of two entangled particles if we neglect the spacetime
dependence of the wave function. 
It was shown in \cite{Vol1} that if one takes into account 
the space part of the
wave function then  the quantum correlation describing
local observations of spins
in the simplest case will take
the form $g \cos (\alpha - \beta)$ instead of just  $\cos
(\alpha - \beta)$. Here the parameter $g$ describes the
location of the system in space and time. In this case 
one  gets a modified equation
$$
g\cos(\alpha - \beta)= E\xi_{\alpha}\eta_{\beta}
$$
One can prove  that if the distance between
detectors is large enough then the factor $g$  becomes small  
and there exists a solution of the modified equation. 
We will show that in fact at large distances
all reasonable quantum states become disentangled.
 This fact leads also to important consequences for 
quantum teleportation and quantum cryptography,  \cite{Vol4,VV2}.
Bell's theorem constitutes an important part in quantum
cryptography.
In \cite{Vol4} it is discussed how one can
try to improve the security of quantum 
cryptography schemes in space  by
using  a special preparation of 
the space part of the wave function.

It is important  to study also a more general question:
which class of functions $f(s,t)$ admits a representation
of the form
$$
f(s,t)=Ex_sy_t
$$
where $x_s$ and $y_t$ are  bounded stochastic processes
and also analogous question for the functions of several variables
$f(t_1,...,t_n).$

Such considerations could provide a {\it noncommutative}
generalization of von Neumann`s spectral theorem.

In this paper entangled states  
in space and time are considered. 
We point out a simple but the fundamental fact
that the vacuum state $\omega_0$ in a free quantum field theory is
a nonfactorized (entangled) state for observables belonging
to spacelike  separated regions:
$$
\omega_0 (\varphi (x)\varphi (y)) - 
\omega_0 (\varphi (x)) \omega_0 (\varphi (y)) 
\neq 0$$
Here $\varphi (x) $ is a free scalar field in the Minkowski
spacetime and $ (x-y)^2 < 0$.
Hence there is a statistical dependence between causally
disconnected regions. 

However one has an asymptotic factorization
of the vacuum state for  large separations of the spacelike 
regions. Moreover one  proves that in quantum field theory
{\it there is 
an asymptotic factorization
for any reasonable state and any local observables. 
Therefore at large distances any reasonable state becomes disentangled}.
We have the relation
$$
\lim_{|l|\to\infty}[\omega(A(l)B)-\omega(A(l))\omega(B)]=0
$$
Here $\omega$ is a state from a rather wide class of the states
which includes entangled states, $A$ and $B$ are two local
observables, and $A(l)$ is the translation of the observable $A$
along the 3 dim vector $l$. As a result a  
violation of Bell`s  inequalities (see below) can be observed
without inconsistency with principles of relativistic quantum theory only
if the distance between detectors is rather small.
We suggest a further experimental study of entangled states 
in spacetime by studying
the dependence of the correlation functions on the distance
between detectors.

There is no a factorization of the expectation value
$\omega_0 (\varphi (x)\varphi (y)) $ even for the space-like
separation of the variables $x$ and $y$ if the distance
between $x$ and $y$ is not large enough. However we will prove that
there exist a representation of the form
$$
\omega_0 (\varphi (x)\varphi (y)) =E\xi (x)\xi^*(y)$$
which is valid for all $x$ and $y$. Here $\xi (x)$
is a classical (generalized) complex  random field and $E$ is the
 expectation value. Therefore the quantum correlation function
is represented as a classical correlation function
of separated random fields. This representation can be called
a local realistic representation
by analogy with the Bell approach to the spin correlation
functions.    

In the next section Bell`s theorem is discussed and
a slight generalization of the known CHSH result
is proved. In Sect.3 the locality in space
is considered for entangled states and
the asymptotic factorization of the states is proved.
A hidden variable representation for quantum
correlation which is local in the space
is also obtained. Noncommutative spectral theory and local realism
are considered in Sect.4. Finally the disentanglement at
large distances in quantum field theory is considered in Sect.5.

\section{Bell's Theorem}

\subsection{Bell`s Theorem and Stochastic Processes}

In the  presentation of Bell's theorem we will 
follow ~\cite{Vol1} where one
can find also more references. Bell's
theorem reads:
\begin{equation}
\cos(\alpha - \beta)\neq E\xi_{\alpha}\eta_{\beta}
\label{eq:cos}
\end{equation}
where $\xi_{\alpha}$ and $\eta_{\beta}$ are two random
processes such that $|\xi_{\alpha}|\leq 1$,~
$|\eta_{\beta}|\leq 1$ and $E$ is the expectation.
In more details:

{\bf Theorem 1.} There exists no probability space 
$(\Lambda, {\cal F}, d\rho (\lambda))$ and a pair of stochastic processes
$\xi_{\alpha}=\xi_{\alpha}(\lambda), ~\eta_{\beta}=\eta_{\beta}(\lambda),
~0\leq \alpha,\beta \leq 2\pi$ which obey 
$|\xi_{\alpha}(\lambda)|\leq 1$,~
$|\eta_{\beta}(\lambda)|\leq 1$ such that the following equation is valid
\begin{equation}
\cos(\alpha - \beta) = E\xi_{\alpha}\eta_{\beta}
\label{eq:cosin}
\end{equation}
for all $\alpha$ and $\beta$.

Here $\Lambda$ is a set, ${\cal F}$ is a sigma-algebra of subsets
and $d\rho (\lambda)$ is a probability measure, i.e. $d\rho (\lambda)
\geq 0,~\int d\rho (\lambda)=1.$
The expectation is
$$
E\xi_{\alpha}\eta_{\beta}=\int_{\Lambda}\xi_{\alpha}(\lambda)
\eta_{\beta}(\lambda)d\rho (\lambda)
$$
One can write Eq.~(\ref{eq:cosin}) as an integral equation
\begin{equation}
\cos(\alpha - \beta)=\int_{\Lambda}\xi_{\alpha}(\lambda)
\eta_{\beta}(\lambda)d\rho (\lambda)
\label{eq:integ}
\end{equation}
We say that the integral equation  (\ref{eq:integ}) has no solutions 
$(\Lambda, {\cal F}, d\rho (\lambda), 
\xi_{\alpha}, \eta_{\beta})$ with the bound
$|\xi_{\alpha}|\leq 1$,~
$|\eta_{\beta}|\leq 1.$

We will prove the theorem below. Let us
discuss now the physical interpretation of this result.

Consider a pair of spin one-half particles 
formed in the singlet spin state
and moving freely towards two detectors.  If one neglects
the space part of the wave function  then one has 
the Hilbert space $C^2\otimes C^2$ 
and  the quantum mechanical
correlation of two spins in the singlet state $\psi_{spin}\in
C^2\otimes C^2$ is
\begin{equation}
 D_{spin}(a,b)=\left<\psi_{spin}|\sigma\cdot a \otimes\sigma\cdot
b|\psi_{spin}\right>=-a\cdot b
\label{eq:eqn1}
\end{equation}
Here $a=(a_1,a_2,a_3)$ and $b=(b_1,b_2,b_3)$ 
are two unit vectors in three-dimensional space $R^3$,
$\sigma=(\sigma_1,\sigma_2,\sigma_3)$ are the Pauli matrices,
$$
\sigma_1=\left(
    \begin{array}{cc}0 & 1\\ 1 & 0
    \end{array}
\right),~~
\sigma_2=\left(
    \begin{array}{cc}0 & -i\\ i & 0
    \end{array}
\right),~~
\sigma_3=\left(
    \begin{array}{cc}1 & 0\\ 0 & -1
    \end{array}
\right),~~
 \sigma\cdot a =\sum_{i=1}^{3}\sigma_i a_i
 $$
 and
$$
\psi_{spin}=\frac{1}{\sqrt 2}
\left(\left(
    \begin{array}{c}0\\1
    \end{array}
    \right)
\otimes \left(
    \begin{array}{c}1\\
    0\end{array}
    \right)
-\left(
    \begin{array}{c}1\\
    0\end{array}
    \right)
\otimes
\left(
    \begin{array}{c}0\\
    1\end{array}
    \right)
\right)
$$
If the vectors $a$ and $b$ belong to the same plane
then one can write $-a\cdot b=\cos (\alpha - \beta)$
and hence Bell's theorem states that the function $ D_{spin}(a,b)$
Eq.~(\ref{eq:eqn1}) can not be represented in the form
\begin{equation}
\label{eq:eqn2} P(a,b)=\int \xi (a,\lambda) \eta (b,\lambda)
d\rho(\lambda)
\end{equation}
i.e.
\begin{equation}
\label{eq:Ab}
D_{spin}(a,b)\neq P(a,b)
\end{equation}
Here $ \xi (a,\lambda)$ and $  \eta(b,\lambda)$ are random  fields on the
sphere, $|\xi (a,\lambda)|\leq 1$,~  $ | \eta (b,\lambda)|\leq 1$ and
$d\rho(\lambda)$ is a positive probability measure,  $ \int
d\rho(\lambda)=1$. The parameters $\lambda$ are interpreted as hidden
variables in a realist theory. It is clear that Eq.~(\ref{eq:Ab}) can be
reduced to Eq.~(\ref{eq:cos}).

\subsection{CHSH Inequality}

To prove Theorem 1 we will use the following theorem which
is a slightly generalized  Clauser-Horn-Shimony-Holt
(CHSH)  result.

{\bf Theorem 2.} Let $f_1,~f_2,~g_1$ and $g_2$ be random variables
(i.e. measured functions) on the probability space $(\Lambda, {\cal F}, d\rho (\lambda))$
such that 
$$
|f_i(\lambda)g_j(\lambda)|\leq 1,~~i,j=1,2.
$$
Denote
$$
P_{ij}=Ef_ig_j,~~i,j=1,2.
$$
Then
$$
|P_{11}-P_{12}|+|P_{21}+P_{22}|\leq 2.
$$
{\bf Proof of Theorem 2.} One has
$$
P_{11}-P_{12}=Ef_1g_1-Ef_1g_2=E(f_1g_1(1\pm f_2g_2))-
E(f_1g_2(1\pm f_2g_1))
$$
Hence
$$
|P_{11}-P_{12}|\leq E(1\pm f_2g_2)+
E(1\pm f_2g_1)=2\pm (P_{22}+P_{21})
$$
Now let us note that if $x$ and $y$ are two real numbers then
$$
|x|\leq 2\pm y~~\to~~|x|+|y|\leq 2.
$$
Therefore  taking $x=P_{11}-P_{12}$ and $y=P_{22}+P_{21}$
one gets the bound
$$
|P_{11}-P_{12}|+|P_{21}+P_{22}|\leq 2.
$$
The theorem is proved. 

The last inequality is called
  the CHSH inequality.
By using notations of Eq.~(\ref{eq:eqn2}) one has
\begin{equation}
\label{eq:eqn3}
 |P(a,b)-P(a,b')|+|P(a',b)+P(a',b')|\leq 2
\end{equation}
for any four unit vectors $a,b,a',b'$.

{\bf Proof of Theorem 1.} Let us denote
$$
f_i(\lambda)=\xi_{\alpha_i}(\lambda),~~
g_j(\lambda)=\eta_{\beta_j}(\lambda),~~i,j=1,2
$$
for some $\alpha_i,\beta_j.$ If one would have
$$
\cos (\alpha_i - \beta_j)=Ef_ig_j
$$
then due to Theorem 2 one should have
$$
|\cos (\alpha_1 - \beta_1)-\cos (\alpha_1 - \beta_2)|
+|\cos (\alpha_2 - \beta_1)+\cos (\alpha_2 - \beta_2)|\leq 2.
$$
However  for $\alpha_1=\pi/2,~\alpha_2=0,~\beta_1=\pi/4,
\beta_2=-\pi/4$ we obtain
$$
|\cos (\alpha_1 - \beta_1)-\cos (\alpha_1 - \beta_2)|
+|\cos (\alpha_2 - \beta_1)+\cos (\alpha_2 - \beta_2)|=2\sqrt 2
$$
which is greater than 2.
This contradiction proves Theorem 1.

It will be shown below that if one takes 
into account the space part of the
wave function then  the quantum correlation 
in the simplest case will take
the form $g \cos (\alpha - \beta)$ instead of just  $\cos
(\alpha - \beta)$ where the parameter $g$ describes the
location of the system in space and time. In this case one can get a
representation
\begin{equation}
g\cos(\alpha - \beta)= E\xi_{\alpha}\eta_{\beta}
\label{eq:gcos}
\end{equation}
if $g$ is small enough. The factor $g$ gives a contribution to
visibility or efficiency of detectors that are used in the phenomenological
description of detectors.

\section {Local Observations}

\subsection{Modified Bell`s equation}

In the previous section the space part of the wave function of the
particles was neglected. However exactly the space part is relevant to the
discussion of locality. The Hilbert space assigned to
one particle with spin 1/2 is  $C^2\otimes L^2(R^3)$
and the Hilbert space of two particles is
$C^2\otimes L^2(R^3)\otimes 
C^2\otimes L^2(R^3).$
The complete wave function is $\psi
=(\psi_{\alpha\beta}({\bf r}_1,{\bf r}_2,t))$ where $\alpha$ and
$\beta $ are spinor indices, $t$ is time  and ${\bf r}_1$ and ${\bf r}_2$
are vectors in three-dimensional space.

We suppose that there are two  detectors (A and B)
which are located in space $R^3$ within the
two localized regions ${\cal O}_A$ and ${\cal O}_B$ respectively, well
separated from one another. 
If one makes a local observation in the region ${\cal O}_A$ 
then this means that one measures not only the spin
observable $\sigma_i$ but also some another observable
which describes the localization of the particle like
the energy density  or the projection operator 
$P_{{\cal O}}$ to the region ${\cal O}$. We will consider here
corelation functions of the projection operators $P_{{\cal O}}$. 

Quantum correlation describing the localized
measurements of spins in the regions ${\cal O}_A$ and ${\cal O}_B$ 
is

\begin{equation}
\label{eq:eqn6}
\omega(\sigma\cdot a   P_{{\cal O}_A}\otimes  
\sigma\cdot b  P_{{\cal O}_B})=\left<\psi|
\sigma\cdot a   P_{{\cal O}_A}\otimes  \sigma\cdot b  P_{{\cal O}_B} 
|\psi\right>
\end{equation}

Let us consider the simplest 
case when the wave function has the form of the product
of the spin function and the space function $\psi=\psi_{spin}\phi({\bf
r}_1,{\bf r}_2)$. Then one has
\begin{equation}
\label{eq:eqn7}
 \omega(\sigma\cdot a   P_{{\cal O}_A}\otimes  
\sigma\cdot b  P_{{\cal O}_B})=
 =g ({\cal O}_A,{\cal O}_B)
  D_{spin}(a,b)
\end{equation}
where the function
\begin{equation}
\label{eq:eqn8}
 g ({\cal O}_A,{\cal O}_B)=\int_{{\cal O}_A \times {\cal O}_B}|\phi({\bf
 r}_1,{\bf
 r}_2)|^2 d{\bf r}_1d{\bf r}_2
\end{equation}
describes correlation of particles in space. It is the probability to find
one particle in the region ${\cal O}_A$ and another particle in the region
${\cal O}_B$.

One has
\begin{equation}
\label{eq:eqn8g} 0\leq g ({\cal O}_A,{\cal O}_B)\leq 1.
\end{equation}

If ${\cal O}_A$ is a bounded region and ${\cal O}_A(l)$
is a translation of ${\cal O}_A$ to the 3-vector $l$
then one has

\begin{equation}
\label{eq:eqn8l} \lim_{|l|\to\infty} g({\cal O}_A(l),{\cal O}_B)=0.
\end{equation}

Since
$$\left<\psi_{spin}|\sigma\cdot a \otimes
I|\psi_{spin}\right>=0
$$
we have
$$
\omega 
(\sigma\cdot a P_{{\cal O}_A}\otimes I)=0.
$$
Therefore
we have proved the following proposition which says that
the state  $\psi=\psi_{spin}\phi({\bf
r}_1,{\bf r}_2)$
becomes disentangled at large distances.

{\bf Proposition.} One has the following property of the asymptotic
factorization (disentanglement) at large distances:
 \begin{equation}
\label{eq:eqn8ld} \lim_{|l|\to\infty}
[\omega (\sigma\cdot a P_{{\cal O}_A(l)}\otimes 
\sigma\cdot b P_{{\cal O}_B})-
\omega 
(\sigma\cdot a P_{{\cal O}_A(l)}\otimes I
)\omega(I\otimes \sigma\cdot b P_{{\cal O}_B}
)]=0
\end{equation}
or
$$
\lim_{|l|\to\infty}
\omega (\sigma\cdot a P_{{\cal O}_A(l)}\otimes 
\sigma\cdot b P_{{\cal O}_B})=0.
$$
Now one inquires whether one can write a representation
\begin{equation}
\label{eq:eqn9}
 \omega(\sigma\cdot a   P_{{\cal O}_A(l)}\otimes  
\sigma\cdot b  P_{{\cal O}_B})=
 \int \xi (a,{\cal O}_A,\lambda)
 \eta (b,{\cal O}_B,\lambda) d\rho(\lambda)
\end{equation}
where $|\xi (a,{\cal O}_A(l),\lambda)|\leq 1,~~
|\eta (b,{\cal O}_B,\lambda)|\leq 1$.

{\bf Remark.} A local modified equation reads
$$
|\phi ({\bf r_1},{\bf r_2},t)|^2\cos(\alpha - \beta)
=E\xi (\alpha,{\bf r_1},t)
\eta (\beta,{\bf r_2},t).
$$

If we are interested in the conditional probability of
finding the projection of spin along  vector $a$ for the particle
1  in the region ${\cal O}_A(l)$ and the projection of spin along the
vector $b$ for the particle 2 in the region ${\cal O}_B$   then we
have to divide both sides of Eq.~(\ref{eq:eqn9}) by $g({\cal
O}_A(l),{\cal O}_B)$.

Note  that here the classical random variable 
$\xi=\xi (a,{\cal O}_A(l),\lambda)$ is not only separated
in the sense of Bell (i.e. it depends only on $a$) 
but it is also local
in the 3 dim space
since it depends only on the 
region ${\cal O}_A(l)$. The classical random variable $\eta$
is also local 
in 3 dim space since it depends only on ${\cal O}_B$.
Note also that since the eigenvalues of the
projector $P_{{\cal O}}$ are 0 or 1 then one should have
$ |\xi (a,{\cal O}_A)|\leq 1.$

Due to the property of the asymptotic factorization
and the vanishing of the quantum correlation
for large $|l|$ there exists a trivial asymptotic
classical representation of the form (\ref{eq:eqn9})
with $\xi=\eta=0.$

We can do even better and find a classical representation which
will be valid uniformly for large $|l|$.

If $g$ would not depend on ${\cal O}_A$ and ${\cal O}_B$
then
instead of Eq~(\ref{eq:cosin}) in Theorem 1  we could have a modified
equation
\begin{equation}
g\cos(\alpha - \beta) = E\xi_{\alpha}\eta_{\beta}
\label{eq:cosinus}
\end{equation}
The factor $g$ is important. In particular one can write the following
representation \cite{VV} for $0\leq g\leq 1/2$:
\begin{equation}
\label{eq:gek}
g\cos(\alpha-\beta)=
\int_0^{2\pi}\sqrt {2g}\cos(\alpha-\lambda) \sqrt {2g}\cos(\beta-\lambda)
 \frac{d\lambda}{2\pi}
\end{equation}
Therefore if $0\leq g\leq 1/2$ then there exists a solution
of Eq.~(\ref{eq:cosinus}) where
$$
\xi_{\alpha}(\lambda)=\sqrt {2g}\cos(\alpha-\lambda),~
~\eta_{\beta}(\lambda)=\sqrt {2g}\cos(\beta-\lambda)
$$
and $|\xi_{\alpha}|\leq 1,~|\eta_{\beta}|\leq 1.$
If $g>1/\sqrt 2$ then it follows from Theorem 2 that there is no
solution to Eq.~(\ref{eq:cosinus}). We have obtained

{\bf Theorem 3.} If $g>1/\sqrt 2$ then there is no
solution 
$(\Lambda, {\cal F}, d\rho (\lambda), 
\xi_{\alpha}, \eta_{\beta})$ to Eq.~(\ref{eq:cosinus}) with the bounds
$|\xi_{\alpha}|\leq 1$,~
$|\eta_{\beta}|\leq 1.$ If $0\leq g\leq 1/2$ then there exists a solution
to Eq.~(\ref{eq:cosinus}) with the bounds
$|\xi_{\alpha}|\leq 1$,~
$|\eta_{\beta}|\leq 1.$

{\bf Remark.} Further results on solutions of the modified
equation have been obtained by A.K. Guschchin,
S. V. Bochkarev and D. Prokhorenko.
Local variable models for inefficient detectors are presented in
\cite{San,Lar}.

Let us take now the wave
function $\phi$ of the form $\phi=\psi_{1}({\bf r}_1)\psi_{2}({\bf
r}_2)$  where
$$
\int_{R^3}|\psi_{1}({\bf r}_1)|^2d{\bf r}_1=1,~~
\int_{R^3}|\psi_{2}({\bf r}_2)|^2d{\bf r}_2=1
$$
In this case
$$
 g ({\cal O}_A(l),{\cal O}_B)=
 \int_{{\cal O}_A(l)}|\psi_{1}({\bf r}_1)|^2d{\bf r}_1\cdot
 \int_{{\cal O}_B}|\psi_{2}({\bf r}_2)|^2d{\bf r}_2
$$
There exists such $L>0$ that
$$
\int_{B_L}|\psi_{1}({\bf r}_1)|^2d{\bf r}_1=\epsilon <1/2,~~
$$
where $B_L=\{{\bf r}\in R^3: |{\bf r}|\geq L\}.$
Let us  make an additional assumption that 
the classical random variable  has the form
of a product of two independent classical random variables
$ \xi (a,{\cal O}_A)=\xi_{space}({\cal O}_A)\xi_{spin}(a)$
and similarly for $\eta.$
We have the following 

{\bf Theorem 4.}
Under the above assumptions and for large enough $|l|$
there exists the following representation of the quantum correlation
function
$$
 g ({\cal O}_A(l),{\cal O}_B)\cos(\alpha-\beta)
 =(E\xi_{space}({\cal O}_A)(l))(E\eta_{space}({\cal O}_B))
 E\xi_{spin}(\alpha)\xi_{spin}(\beta)
 $$
 where all classical random variables are bounded by 1.
 
{\bf Proof.} To prove the theorem we write
$$
g ({\cal O}_A(l),{\cal O}_B)\cos(\alpha-\beta)
=\int_{{\cal O}_A(l)}\frac{1}{\epsilon}|\psi_{1}({\bf r}_1)|^2d{\bf r}_1
\cdot
 \int_{{\cal O}_B}|\psi_{2}({\bf r}_2)|^2d{\bf r}_2
\cdot \epsilon \cos (\alpha-\beta)
$$
$$=(E\xi_{space}({\cal O}_A(l))(E\eta_{space}({\cal O}_B))
 E\xi_{spin}(\alpha)\xi_{spin}(\beta)
$$
Here  $\xi_{space}({\cal O}_A(l))$ and
$\eta_{space}({\cal O}_B)$ are random variables on the probability space
$B_L\times R^3$ with the probability measure
$$
dP({\bf r}_1,{\bf r}_2)=
\frac{1}{\epsilon}|\psi_{1}({\bf r}_1)|^2
\cdot
 |\psi_{2}({\bf r}_2)|^2d{\bf r}_1d{\bf r}_2
$$
of the form 
$$
\xi_{space}({\cal O}_A(l),{\bf r}_1,{\bf r}_2)
=\chi_{{\cal O}_A(l)}({\bf r}_1),~~
\eta_{space}({\cal O}_B,{\bf r}_1,{\bf r}_2)
=\chi_{{\cal O}_B}({\bf r}_2)$$
where $\chi_{{\cal O}}({\bf r})$ is the characteristic
function of the region ${\cal O}.$ We assume that
${\cal O}_A(l)$ belongs to $B_L.$
Further $\xi_{spin}(\alpha)$ is a random process
on the circle $0\leq \varphi \leq 2\pi$ with the probability
measure $d\varphi /2\pi$ of the form
$$
\xi_{spin}(\alpha,\varphi)=\sqrt{2\epsilon}\cos(\alpha - \varphi)
$$
The theorem is proved.
 
\subsection{Expansion of Wave Packet}

Let us remind that there is a well known effect of expansion of
wave packets due to the free time evolution. If $\epsilon$ is the
characteristic length of the Gaussian wave packet describing a
particle of mass $M$ at time $t=0$  then at time $t$ the
characteristic length  $\epsilon_t$ will be
\begin{equation}
\label{eq:epsil}
\epsilon_t=\epsilon\sqrt{1+\frac{\hbar^2t^2}{M^2\epsilon^4}}.
\end{equation}

It tends to $(\hbar/M\epsilon)t$ as $t\to\infty$. Therefore the
locality criterion is always satisfied for nonrelativistic
particles if regions ${\cal O}_A$ and ${\cal O}_B$ are far enough
from each other.

 \subsection {Relativistic Particles}
 
We can not immediately apply the previous considerations
to the case of relativistic particles such as photons
and the Dirac particles because in these cases
 the wave function can not be represented 
as a product of the spin part and the spacetime part.
Let us show that the wave function of photon can not
be represented in the product form. Let $A_i(k)$ be the wave
function of photon, where $i=1,2,3$ and $k\in R^3$.
One has the gauge condition $k^iA_i(k)=0$ \cite{AB}. If one supposes
that the wave function has a product form $A_i(k)=\phi_i f(k)$
then from the gauge condition one gets $A_i(k)=0$.
Therefore the case of relativistic particles requires
a separate investigation (see below). 

\section{Noncommutative Spectral Theory and Local Realism}

As a generalisation of the previous discussion
we would like to suggest here
a general relation between quantum theory
and  theory of classical stochastic processes
 which expresses the condition of local realism.
Let ${\cal H}$ be a Hilbert space, $\rho $ is the density
operator, $\{A_{\alpha}\}$ is a family of self-adjoint
operators in ${\cal H}$. One says that the family
of observables $\{A_{\alpha}\}$ and the state $ \rho$
satisfy to {\it the condition of local realism} if there exists
a probability space $(\Lambda, {\cal F}, d\rho (\lambda))$
and a family of random variables $\{\xi_{\alpha}\}$ such
that  the range of $\xi_{\alpha}$ belongs to the spectrum
of $A_{\alpha}$ and for
any subset $\{A_i\}$ of mutually commutative operators
one has a representation
$$
Tr( \rho A_{i_1}...A_{i_n})=E\xi_{i_1}...\xi_{i_n}
$$  
The physical meaning of the representation is that it 
describes the quantum-classical correspondence.
If the family $\{A_{\alpha}\}$ would be a maximal commutative
family of self-adjoint operators then for pure states
the previous representation
can be reduced to the von Neumann spectral
theorem~\cite{Nai}. In our case the family $\{A_{\alpha}\}$ 
consists  from not necessary commuting operators.
Hence we will call such a representation a {\it noncommutative spectral 
representation}.  Of course one has a question for which families
of operators and states a {\it noncommutative
spectral theorem}  is valid, i.e. when we can write 
the noncommutative spectral representation. We need a noncommutative
generalization of von Neumann`s spectral theorem.

It would be helpful
to study the following problem: describe the class of
functions $f(t_1,...,t_n)$ which admits
the representation of the form
$$
f(t_1,...,t_n)=Ex_{t_1}...z_{t_n}
$$
where $x_t,...,z_t$ are random processes which obey the bounds
$|x_t|\leq 1,...,|z_t|\leq 1$. 

From the previous discussion we know that there are such 
families of operators and such states which do not 
admit the noncommutative spectral representation
and therefore they do not satisfy
the condition of local realism. Indeed let us take
the Hilbert space ${\cal H}=C^2\otimes C^2$
and four operators $A_1,A_2,A_3,A_4$ of the form (we denote
$A_3=B_1, A_4=B_2$)
$$
A_1=\left(
    \begin{array}{cc}\sin\alpha_1 & \cos\alpha_1
	\\ \cos\alpha_1 & -\sin\alpha_1
    \end{array}
\right)\otimes I,~~A_2=\left(
    \begin{array}{cc}\sin\alpha_2 & \cos\alpha_2
	\\ \cos\alpha_2 & -\sin\alpha_2
    \end{array}
\right)\otimes I
$$
and
$$
B_1=I\otimes\left(
    \begin{array}{cc}-\sin\beta_1 & -\cos\beta_1
	\\ -\cos\beta_1 & \sin\beta_1
    \end{array}
\right),~~B_2=I\otimes\left(
    \begin{array}{cc}-\sin\beta_2 & -\cos\beta_2
	\\ -\cos\beta_2 & \sin\beta_2
    \end{array}\right)
$$
Here operators $A_i$ correspond to operators $\sigma\cdot a$
and operators $B_i$ corresponds to operators $\sigma\cdot b$
where $a=(\cos\alpha,0,\sin\alpha),~b=(-\cos\beta,0,-\sin\beta).$
Operators $A_i$ commute with operators 
$B_j,~~[A_i,B_j]=0,~i,j=1,2$ and one has
$$
\left<\psi_{spin}|A_iB_j|\psi_{spin}\right>
=\cos(\alpha_i - \beta_j),~~i,j=1,2
$$
We know from Theorem 2 that this function can not be represented
as the expected value $E\xi_i\eta_j$  of random variables
with the bounds $|\xi_{i}|\leq 1$,~
$|\eta_{j}|\leq 1.$
 
However, as it was discussed above, the space part of the wave
function
was neglected in the previous consideration.
We suggest that {\it in physics one could prepare
only such states and observables which satisfy the condition
of local realism.} Perhaps we should restrict ourself
in this proposal to the consideration of  
only such families of observables
which satisfy the condition of relativistic local causality.
If there are physical phenomena which do not satisfy this proposal
then it would be important to {\it describe quantum processes
which satisfy the above formulated condition of local realism 
and also processes
which do not satisfy this condition}.

\section{Quantum Probability and Quantum Field Theory}

In quantum probability (see \cite{ALV})
we are given a * - algebra ${\cal A}$
and a state (i.e. a linear positive normalized
functional) $\omega$ on ${\cal A}$. Elements from ${\cal A}$
are called random variables. Two random variables $A$ and $B$
are called (statistically) independent 
if $ \omega(AB)=\omega(A)\omega(B)$. 

First we will prove the following

{\bf Proposition}. {\it There is a statistical 
dependence between
two spacelike separated regions in the 
theory of free scalar quantum field}.

{\bf Proof}.
Let us consider a free scalar quantum field $\varphi (x)$:
$$
\varphi (x) =\frac{1}{(2\pi)^{3/2}}\int_{R^3}
\frac{d{\bf k}}{\sqrt{2k^0}}(e^{ikx}a^*({\bf k})+
e^{-ikx}a({\bf k}))
$$
Here $kx=k^0x^0-{\bf kx}$, $k^0=\sqrt{{\bf k}^2+m^2}, m\leq 0$
and $a({\bf k})$ and $a^*({\bf k})$ are  annihilation and creation
operators,
$$
[a({\bf k}), a^*({\bf k'})]=\delta ({\bf k}-{\bf k'})
$$
The field $\varphi (x)$
is an operator valued distribution acting in the Fock space
${\cal F}$ with the vacuum $|0>$,
$$
a({\bf k})|0>=0
$$
The vacuum expectation value of two fields is
$$
\omega_0 (\varphi (x)\varphi (y))=<0|\varphi (x)\varphi (y)|0>
=W_0(x-y, m^2)
$$
where
$$
W_0(x-y, m^2)=\frac{1}{(2\pi)^{3}}\int_{R^3}
\frac{d{\bf k}}{2k^0}e^{-ik(x-y)}
$$
The statistical independence of two spacelike separated regions in
particular would lead to the relation
$$
\omega_0 (\varphi (x)\varphi (y)) - 
\omega_0 (\varphi (x)) \omega_0 (\varphi (y)) 
= 0
$$
if $(x-y)^2<0$. But since $\omega_0 (\varphi (x))=0$
in fact we have
$$
\omega_0 (\varphi (x)\varphi (y)) - 
\omega_0 (\varphi (x)) \omega_0 (\varphi (y)) 
= W_0(x-y, m^2)\neq 0
$$
However the violation of
the statistical independence vanish exponentially with the spacial
separation of $x$ and $y$ since for large $\lambda =m\sqrt {-x^2}$
the function $W_0(x, m^2)$ behaves like
$$
\frac{m^2}{4\pi \lambda}\left(\frac{\pi}
{2\lambda}\right)^{1/2}e^{-\lambda}
$$

Let us prove that any polynomial state is asymptotically
disentangled (factorized) for large spacelike distances.
Let ${\cal A}$ be the algebra of polinomials 
in the Fock space
${\cal F}$
at the field
$\varphi (f)$ with the test functions $f$ .
Let $C\in {\cal A}$ and $|\psi>=C|0>.$
Denote the state $\omega (A)=\left<\psi|A|\psi\right >/||\psi||^2$
for $A\in {\cal A}$.

{\bf Theorem 5.} One has the following asymptotic
disentanglement property
$$
\lim_{|l|\to\infty}[\omega(A(l)B)-\omega(A(l))\omega(B)]=0
$$
Here  $A$ and $B$ belong to ${\cal A}$ 
and $A(l)$ is the translation of $A$
along the 3 dim vector $l$. One has also
$$
\lim_{|l|\to\infty}[\omega(A(l))-\left <0|A(l)|0\right >]=0
$$
The proof of the theorem is based on the Wick theorem
and the Riemann-Lebesgue lemma.

Similar theorems take place also for the Dirac and the Maxwell
fields. In particular for the Dirac field $\psi (x)$
one can prove the asymtotic factorization for the local spin operator
$$
{\bf S}({\cal O})=\int_{{\cal O}}\psi^*{\bf \Sigma}\psi dx
$$
Here $\Sigma$ is made from the Dirac matrices.

Finally let us show that some correlation functions
in the relativistic quantum field theory can be
represented as mathematical expectations of
the classical (generalized) random fields.

{\bf Theorem 6.} If $\varphi (x)$ is a scalar complex quantum field
then one has a representation
$$
\left <0|\varphi (x_1)...\varphi (x_n)\varphi^* (y_1)
...\varphi^* (y_n)|0\right >=E\xi (x_1)...\xi (x_n)\xi^* (y_1)
...\xi^* (y_n).
$$
Here $\xi (x)$ is a complex random field.

The proof of the theorem follows from 
the positivity of the quantum correlation
functions. It is interesting that we have obtained a 
functional integral representation for the quantum correlation functions
in real time. Similar representation is valid also for 
the 2-point correlation function of an interacting scalar
field. It follows from the Kallen-Lehmann representation.

\section {Conclusions}

We have discussed some problems in quantum information theory
which requires the inclusion of spacetime variables.
In particular entangled states  
in space and time were considered. A modification of
Bell`s equation which includes the spacetime variables is suggested and 
investigated.
A general relation between quantum theory
and  theory of classical stochastic processes was proposed
 which expresses the condition of local realism
 in the form of a noncommutative spectral theorem.
  Applications of this relation to the security 
of quantum key distribution
in quantum cryptography was mentioned.

There are many interesting open problems in the approach
to quantum information in space and time discussed in this paper.
Some of them related with the noncommutative spectral theory and
theory of classical stochastic processes have been discussed above.
It would be useful if the local algebraic approach 
to quantum theory \cite{Bog} will be developed in this direction.

 Entangled states   in space and time are
considered. It is  shown that any reasonable state in  relativistic quantum
field theory  becomes disentangled (factorizable) at large spacelike
distances if one makes local observations. 
As a result a  violation of Bell`s  inequalities can be observed
without inconsistency with principles of relativistic quantum theory only
if the distance between detectors is rather small.
We suggest a further experimental study of entangled states 
in spacetime by studying
the dependence of the correlation functions on the distance
between detectors.

\section*{Acknowledgments}

I am grateful to R. Gill,  
A. Holevo, 
A. Khrennikov, T.Ishiwatari, S. Lloyd, S. Molotkov,  M. Ohya, 
R. Roschin, A. Suarez
 and K.A. Valiev
for helpful discussions and remarks.
This work  is supported in part by RFFI 
 and by INTAS 99-00545 grants.

\newpage

\end{document}